\documentclass[12pt,preprint]{aastex}
\begin{document}
\title{Beryllium in the Ultra-Lithium-Deficient, \\
Metal-Poor Halo Dwarf, G186-26\altaffilmark{1}}

\author{Ann Merchant Boesgaard\altaffilmark{2} and Megan C. Novicki\altaffilmark{2}}

\altaffiltext{1}{Based on data obtained with the Subaru Telescope, which is
operated by the National Astronomical Observatory of Japan.}

\altaffiltext{2}{Institute for Astronomy, University of Hawai`i at M\-anoa, 
2680 Woodlawn Drive, Honolulu, HI {\ \ }96822, boes@ifa.hawaii.edu, 
mnovicki@ifa.hawaii.edu}

\begin{abstract}
The vast majority of low-metal halo dwarfs show a similar amount of Li; this
has been attributed to the Li that was produced in the Big Bang.  However,
there are nine known halo stars with T $>$ 5900 K and [Fe/H] $<$ $-$1.0 that
are ultra-Li-deficient.  We have looked for Be in the very low metallicity
star, G 186-26 at [Fe/H] = $-$2.71, which is one of the ultra-Li-deficient
stars.  This star is also ultra-Be deficient.  Relative to Be in the Li-normal
stars at [Fe/H] = $-$2.7, G 182-26 is down in Be by more than 0.8 dex.  Of two
potential causes for the Li-deficiency -- mass-transfer in a pre-blue
straggler or extra rotationally-induced mixing in a star that was initially a
very rapid rotator -- the absence of Be favors the blue-straggler hypothesis,
but the rotation model cannot be ruled-out completely.
\end{abstract}

\keywords{stars: abundances; stars: evolution; stars: late-type;
subdwarfs; stars: individual (G 186-26); stars: Population II; Galaxy: halo}

\section{Introduction}

Although nearly all metal-poor dwarf and turn-off stars with T$_{\rm eff}$
$>$5600 K and [Fe/H] $\lesssim$$-$1.3 show similar Li abundances, there are a
few such stars which seem to be ultra-deficient in Li.  The discovery by Spite
\& Spite (1982) of a plateau in the Li abundances in metal-poor stars has been
followed by much research to determine Li in many additional ancient halo
stars and to derive the primordial Li abundance produced by Big Bang
nucleosynthesis (BBN) (e.g. Thorburn 1994, Bonifacio \& Molaro 1997, Ryan et
al.~1999, 2000, Mel\'endez \& Ram\'\i rez 2004).  In fact, in the follow-up
work to the initial paper by Spite \& Spite (1982), Spite et al.~(1984)
discovered a Li-deficient star, HD 97916, with a Li abundance at least an
order of magnitude below the plateau.  The plateau value of A(Li) ( = log
N(Li/H) + 12.00) is near 2.2, but ``the plateau'' seems to have both a
temperature and a metallicity dependence, e.g. Novicki (2005); those results
are consistent with the results of Thorburn (1994) and Ryan et al.~(2001a) in
terms of both the coefficients of the dependencies and the sample size.  For
HD 97916 at 6124 K the upper limit on A(Li) is $<$1.2.  The extreme halo star,
G 186-26, was found to be Li deficient by Hobbs, Welty, \& Thorburn (1991).
Additional Li-deficient metal-poor stars have been discovered by Hobbs \&
Mathieu (1991), Thorburn (1994), and Ryan et al.~(2001b).  There are now nine
known halo dwarfs with [Fe/H] $<$$-$1.1 all having T$_{\rm eff}$ $>$ 5980 K.

These ``ultra-Li-deficient stars'' are important in estimating the primordial
Li value, A(Li)$_p$.  Do they represent a true dispersion in the plateau?  Do
they indicate greater Li depletion and are thus sign-posts of general (more
mild) Li depletion in all or most of the plateau stars?  If all plateau stars
have undergone some depletion, then A(Li)$_p$ is higher than the currently
measured value, with concomitant implications for cosmology, such as the
baryon-to-photon ratio, $\eta$.

There are at least two possible origins for the Li deficiencies.  Ryan et
al.~(2001a, 2002) argue that the original Li is reduced by the mechanism that
operates in blue-straggler stars, primarily a mass transfer event or a stellar
merger, but in ``blue-stragglers-to-be.''  Pinsonneault et al.~(1999, 2002)
propose that stellar mixing caused by rotation has lowered Li in the plateau
and has produced larger Li depletions in some fraction of the stars that were
originally the most rapid rotators.  (Ryan et al.~2001a rule out diffusion and
a Hyades-like Li-dip as explanations for the severely Li-deficient halo
stars.)  These two hypotheses have different implications for the Be content
of the ultra-Li-deficient stars.  In the blue-straggler model the star would
undergo substantial internal mixing and/or mass transfer which would destroy
both Li and Be completely as atoms of both elements would be in environments
where the temperatures are high enough to destroy them: $\sim$2.5 x 10$^6$ K
for Li and $\sim$3.5 x 10$^6$ K for Be.  (Mass transferred from an evolved
giant would also be diluted in Li and Be.)  In the rotationally-induced mixing
model Li may be partly or completely destroyed, while Be may be totally or
partly preserved (Pinsonneault, Deliyannis \& Demarque 1992, hereafter PDD92).

The nine ultra-Li-deficient are not a uniform group.  Four of them are
single-line spectroscopic binaries (Carney et al.~2001); three of those four
plus one other have measureable rotation velocities, while two (including G
186-26) have very sharp lines with only upper limits on v sin i (Elliott \&
Ryan 2005); three or four (including G 186-26) are single stars (Ryan et
al.~2001a).  There seem to be no peculiarities in composition that
characterize all of them (e.g. Spite et al.~1993, Norris et al.~1997; Elliott
\& Ryan 2005).

In this letter we report on the Be abundance of one of the warmest and most
metal-poor of the ultra-Li deficient stars, G 186-26.  The value of A(Li) for
this star is $<$1.1 and [Fe/H] = $-$2.8 (Thorburn 1994).

\section{Observations and Analysis}

The spectra for G 186-26 were obtained on one night in 2003 May 27 (UT) with
the high-dispersion spectrograph (HDS) at the Subaru 8.2 m telescope on Mauna
Kea (Noguchi et al.~2002).  We used the blue collimator and blue
cross-disperser with the standard HDS setup: StdUb.  Our slit was 0.7 x 4.4
arcsec and the binning was 2 x 2.  There are two EEV-CCDs covering 2048 x 4100
pixels (pixel size = 13.5 $\mu$) corresponding to a wavelength coverage of
2970 - 4640 \AA.  Calibration exposures were taken of the bias, halogen lamps
for flat-fielding, and Th-Ar comparison spectra.  We obtained two integrations
of 50 minutes each of this V = 10.82 star for a total signal-to-noise (per
pixel) of 98 at 3130 \AA.  The spectral resolution is $\sim$50,000 or 0.062
\AA.  The dispersion is 0.0187 \AA\ pix$^{-1}$ and the resolution element is
3.3 pix.  Both resonance lines of Be II at 3130.416 and 3131.064 \AA\ appear
in two different orders of the spectrum and we analyzed both separately, but
the flux in the order where the Be II lines are centered is 8.5 times that in
the lower order.  Standard data reduction procedures were used.  After
division by the normalized flat field, scattered light and cosmic ray events
were removed.  The spectra were traced and extracted, and wavelength
calibrated from the Th-Ar spectra.  The two exposures were combined and the
continuum was determined.  Since G 186-26 has a metallicity of 500 times below
solar, there is little blending and the continuum placement was
straight-forward.

The stellar parameters for G 186-26 have been determined by Novicki (2005).
We used the photometric indices of $(b-y)$$_0$, $(V-K)$$_0$, and $(R-I)$$_0$
and the calibrations of Carney (1983) to determine the effective temperature,
T$_{\rm eff}$.  (The $(b-y)$ values are from Shuster \& Nissen (1988), $(V-K)$
from Alonso et al.~(1994) and Carney (1983), the $(R-I)$ from Eggen (1979).)
A weighting of 4:2:1 was used for T$_{\rm eff}$($b-y)_0$: T$_{\rm
eff}$($V-K)_0$: T$_{\rm eff}$($R-I)_0$.  (For G 186-26 the reddening is
negligible and was ignored: E($b-y$) = 0.011.)  The three temperatures are
6196 K, 6181 K, and 6256 K with a weighted mean of 6200 K $\pm$25 K, but we
use $\pm$40 K as a more realistic estimate of the uncertainty.  For comparison
Alonso et al.~(1996) find a weighted mean of 6428 K from H, J, K photometry
for this star.  We have found six values for [Fe/H] in the literature from
high-resolution, high S/N determinations.  These have been normalized to the
same scale (e.g. to solar log N(Fe/H) = 7.51) and averaged.  Two lower
resolution results have been included also.  The final weighted mean for
[Fe/H] is $-$2.71 (Novicki 2005).  We used Str\"omgren photometry, ($b-y$)$_0$
and c$_0$, to estimate log g via the Yi, Demarque \& Kim (2004) 10 Gyr
isochrone for Y = 0.23.  This gives a value of 4.48 and we estimate the
uncertainty as 0.20 dex.  (The 13 Gyr isochrone gives log g of 4.44 which
would result in a smaller Be abundance by 0.018 dex, so our value of 4.48 is a
conservative choice.)  The microturbulent velocity was taken to be 1.5 km
s$^{-1}$, the typical value for low-metal stars according to Magain (1987).
Our parameters for G 186-26 are T$_{\rm eff}$ = 6200 $\pm$40 K, log g = 4.48
$\pm$ 0.20, [Fe/H] = $-$2.71 $\pm$0.12, which are in good agreement with those
of Thorburn (1994), Norris et al.~(1997), and Akerman et al.~(2004).

The Be abundance was determined using spectrum synthesis with the program
MOOG, version 2002, (Sneden 1973; http://verdi.as.utexas.edu/moog.html).  The
stellar parameters were used to generate a model atmosphere using the Kurucz
(1993) grid of models.  All elements except Be and O are reduced by the same
amount as [Fe/H].  The O abundance matters as there are many blending OH lines
in the Be II spectral region.  We have used the value of [O/Fe] = +0.49
derived for this star by Akerman et al.~(2004).  Figure 1 shows the spectrum
synthesis fit for Be in our combined spectrum of G 186-26.  Neither of the two
Be II resonance lines is present.  Shown for comparison in Figure 1 is the
Li-normal star, BD +3 740, which has normal Be for its temperature and [Fe/H],
both of which are similar to those parameters for G 186-26; this spectrum is
from Boesgaard et al.~(1999) but has been reanalyzed with the latest version
of MOOG.  Figure 2 shows an enlarged view of the Be II lines in G 186-26 with
the same synthesis as in Figure 1.  The value for A(Be) of $-$1.58 was
selected as the expected initial Be abundance (see $\S$3.2).  The uncertainty
in the Be abundances that is due to the uncertainties in the stellar
parameters is $\pm$0.09.

\section{Results and Interpretation}

It can be seen from Figures 1 and 2 that the spectrum of G 186-26 is
consistent with no Be present in the stellar atmosphere, but we suggest an
upper limit on A(Be) of $-$2.00 $\pm$0.09 dex.  As Figure 1 shows, we have
detected Be II lines in BD +3 740 at 6227 K and [Fe/H] = $-$2.81 and derive
A(Be) = $-$1.37.  Furthermore, Primas et al.~(2000a) show the presence of Be
in G 64-12, a star with a similar temperature (6400 K) and even lower
metallicity, [Fe/H] = $-$3.28, with A(Be) = $-$1.15.

We compare our upper limit on Be for G 186-26 with Be abundances for other
stars at its metallicity.  Our Figure 3 shows the Be abundance limit of G
186-26 in the context of Be detections in other metal-poor, Li-normal stars as
a function of metallicity.  The stars below [Fe/H] of $-$0.9 appear to fall on
the same Fe-Be trend considering the scatter in the Be abundances at a given
Fe, i.e.~they have normal Be abundances.  They also all have normal Li plateau
abundances.  Our A(Be) limit, $<$$-$2.0, is less than other such stars at this
metallicity by 0.8 dex.  Allowing for the $\pm$0.12 uncertainty in [Fe/H], it
is lower by 0.7 - 0.9 dex.

\subsection{Blue Straggler}

Ryan et al.~(2001) pose a question about the blue straggler phenomenon which
essentially is: why should the processes that form blue stragglers be limited
only to those stars that have masses {\it above} the turn-off for main
sequence stars?  Those higher mass stars are apparent because of their unusual
colors -- too bright and too blue for their ages.  The phenomenon involving
mass transfer and mergers could be at work in lower-mass, sub-turnoff stars as
well.  Peculiar colors would not be the signature of this behavior, but Li
depletion would result.  For example, blue stragglers examined for Li in M 67
by Pritchet \& Glaspey (1991) have no detectable Li.  Hobbs \& Mathieu (1991)
found no Li line in blue stragglers in the two halo field stars they studied.
Those two stars were ``known'' blue stragglers before they were on the list
of the nine ultra-Li deficient stars.  Carney et al.~(2005) find that 4 of the
5 Li-deficient field blue stragglers that they studied are single-lined
spectroscopic binaries.

Pritchet \& Glaspey (1991) conclude that the blue straggler formation must be
due to binary coalescence, binary mass transfer, or other deep mixing.  Such
stars would be thoroughly mixed. All three light elements, Li, Be, and B,
would be destroyed in this scenario as all the atoms of these elements would
be subjected to temperatures where they would be destroyed by nuclear
reactions.  Preston \& Sneden (2000) argue that most field blue stragglers are
produced by mass transfer events because collisions are less common in the
field than in clusters.  The former-giant primary that transfers the mass
would deposit Li-depleted deep envelope material onto the current primary.
Our measurement of A(Be) and Thorburn's (1994) measurement of A(Li) show that
both Li and Be are severely depleted -- if present at all.  We may be
observing a star that is a blue-straggler-to-be.  It does not have the mass to
have evolved beyond the main sequence, yet it has the signature of a coalesced
binary or mass transfer in its Li and Be deficiencies.

Latham et al.~(2002) did not find a period for G 186-26.  The Carney et
al.~(1996) radial velocity measurements show a variation from $-$317.78 to
$-$323.19 km s$^{-1}$ for a range of 5.41 km s$^{-1}$ with measurement errors
of $\pm$0.71 to $\pm$1.89 (or $\pm$0.98, the mean internal error).  Their 43
measurements with error bars over 3220 days are shown in Figure 4.  Four
additional measurements by Latham et al.~(2002) some 3300 days later spanning
139 days center around $-$322.6 km s$^{-1}$ with errors of $\sim$0.9 km
s$^{-1}$.  There are no signs of double lines in our spectra.

If there has been mass transfer from a former AGB star onto G 186-26, it would
have left composition signatures.  In particular, Sneden, Preston \& Cowan
(2003) have shown that blue stragglers would have overabundances of the
s-process elements, Sr, Ba, and Pb.  Norris et al.~(1997) found an
overabundance of Ba in G 186-26, with [Ba/Fe] = 0.35.  Abundances of many
elements in six ultra-Li-deficient halo stars were reported by Elliott \& Ryan
(2005); they find that Sr is overabundant by +0.5 dex and Ba by +1.0 dex in G
186-26.  Norris et al.~(2001), Stephens \& Boesgaard (2002) and Fulbright
(2002) all find that such an overabundance of [Ba/Fe] is extremely rare at
this low [Fe/H].

Support for the blue straggler model for the ultra Li-deficient stars comes
from the apparent lack of Be in G 186-26.  Additional support for this comes
from the super-solar content of neutron-capture elements.  However, there is
no evidence that G 186-26 is a binary star, but it may be a coalesced binary.

\subsection{Rotation}

Although all the Li abundances in the ultra Li-deficient stars are upper
limits, the presence of Be could indicate that rotationally-induced mixing has
occurred, but not down to internal temperatures as high as $\sim$3.5 x 10$^6$
K.  Therefore, abundances of Be in the ultra-deficient Li halo stars could
provide a clue about the efficiency of internal mixing.  That is, if Be is
found to be present in the ultra-Li deficient stars, that would imply that the
{\it Li depletion} is likely to be due to mixing caused by high initial
angular momentum.

The effect of rotationally-induced mixing on the Li and Be abundances in halo
stars was first addressed by PDD92.  According to the standard models, there
is no Be depletion (their Tables 3A and 3B).  The models which track the
instabilities due to spin-down and angular momentum loss do predict depletion
of both Li and Be.  These authors address whether there is a range in initial
rotation rates (i.e. in initial angular momentum) that results in a range in
Li at the present epoch.  Since Li is destroyed at shallower layers than Be,
this mixing mechanism should produce less Be destruction than Li destruction,
unless mixing is deep enough to deplete both elements substantially.

We have tried to determine the amount of Li and Be depletion that is predicted
for our star by the PDD92 models.  Therefore, we have examined the latest
``Yale'' isochrones (Yi et al.~2004) appropriate for our star's [Fe/H], log g,
and effective temperature to find the mass for G 186-26.  The 13 Gyr isochrone
gives a mass of 0.735 M$_{\odot}$.  Tables 5A, 5B, and 5C in PDD92 show the
expected depletion for Li and Be for Z = 0.0001 for three values of the
initial angular momentum.  We looked at Table 5C primarily, in order to find
the largest Li and Be depletions.  From our interpolation we find Li depleted
from the initial value by $-$1.21 dex and Be by $-$0.40 dex, compared with at
least $-$0.8 dex in G186-26.  If the initial Li is 2.2 dex (like the current
plateau value), the present value would be 1.0 dex; the measured limit on
A(Li) by Thorburn (1994) is consistent at $<$1.1 dex.  The initial value for
Be can be found from the relationship between A(Be) and [Fe/H] for the iron
abundance in G 186-26 of [Fe/H] = $-$2.71.  From Figure 5 of Boesgaard et
al.~(1999) and Figure 10.3 of Boesgaard (2004) we find the initial Be was
about $-$1.18 dex and thus the current value would be $-$1.18 + $-$0.40 =
$-$1.58 for A(Be).  Figures 1 and 2 show this value in the spectrum synthesis
as well as 2 and 4 times more Be than that, $-$2.00 (2.6 times less Be), and
zero Be.  Although no Be gives the best fit, an upper limit of $-$2.0 dex is
consistent with the noise seen in the continuum and the overall fit of the
synthesis.  We note there are uncertainties in the initial Be assumed for this
star and for the amount of Be depletion from the models.

Newer models and observations on this issue have been discussed by
Pinsonneault et al.~(1999, 2002).  Pinsonneault et al.~(2002) explain that the
early models of PDD92 did not take into account that there could be a
saturation effect in the loss of angular momentum for the rapidly rotating
stars.  They state that the amount of Li depletion could not be {\it as large
as} the PDD92 rates.  This would likely reduce the amount of Be depletion
also.  The hypothesis that the ultra Li-deficient stars were originally part
of a group of very rapidly rotating stars which have now spun down, along with
some destruction of Li and Be, is not easy to reconcile with such a low upper
limit on A(Be).  This issue of Be abundances should be readdressed with modern
models.

\section{Summary and Conclusions}

The nature of the ultra Li-deficient metal-poor stars has been the subject of
several papers and is relevant for the determination of the value of
primordial Li.  Two leading hypotheses for the Li deficiencies are: (1) They
are analogs to blue stragglers which have lost surface Li through binary
mass-transfer or mergers.  (2) They are the descendents of a subset of stars
with initial rapid rotation that have depleted Li by rotationally-induced
mixing during spin-down through their evolution.  In the blue straggler idea
there would be complete destruction of both Li and Be.  If extra mixing
resulted from rotation, much of the Li would be destroyed, but little or no
Be.  Thus Be acts as a discriminator between these two hypotheses because it
is less fragile than Li.
 
We have made observations of Be in a Li-deficient, very metal-poor halo star,
G 186-26.  This star has [Fe/H] = $-$2.71, T$_{\rm eff}$ = 6200, and A(Li)
$<-$1.1.  Our Subaru/HDS spectrum has a spectral resolution of $\sim$50,000
and S/N of 98 in the Be II spectral region near 3130 \AA.  An upper limit for
A(Be) is $<-$2.00.  The data support the blue straggler hypothesis over
rotation.  A coalesced binary and mass transfer episodes could result in the
observed of deficits of Li and Be.

For G 186-26 the blue straggler analog fits well, but it is necessary to look
at Be in other ultra Li-deficient, low-metal stars.  To more fully evaluate
the rotation hypothesis, new model predictions should be done for Be.  Elliott
\& Ryan (2005) have shown that there are no commonalities in other abundances,
but four of their six stars have measureable rotational velocities; perhaps
rotation plays a role in these stars.

\acknowledgements We thank Dr. Sean Ryan for his suggestions on an earlier
version of this paper.  We are grateful for the assistance at the telescope by
the enthusiastic crew at Subaru.  This work was supported by NSF grants
AST-0097945 and AST-0505899 to A.M.B.

\begin{figure} 
\plotone{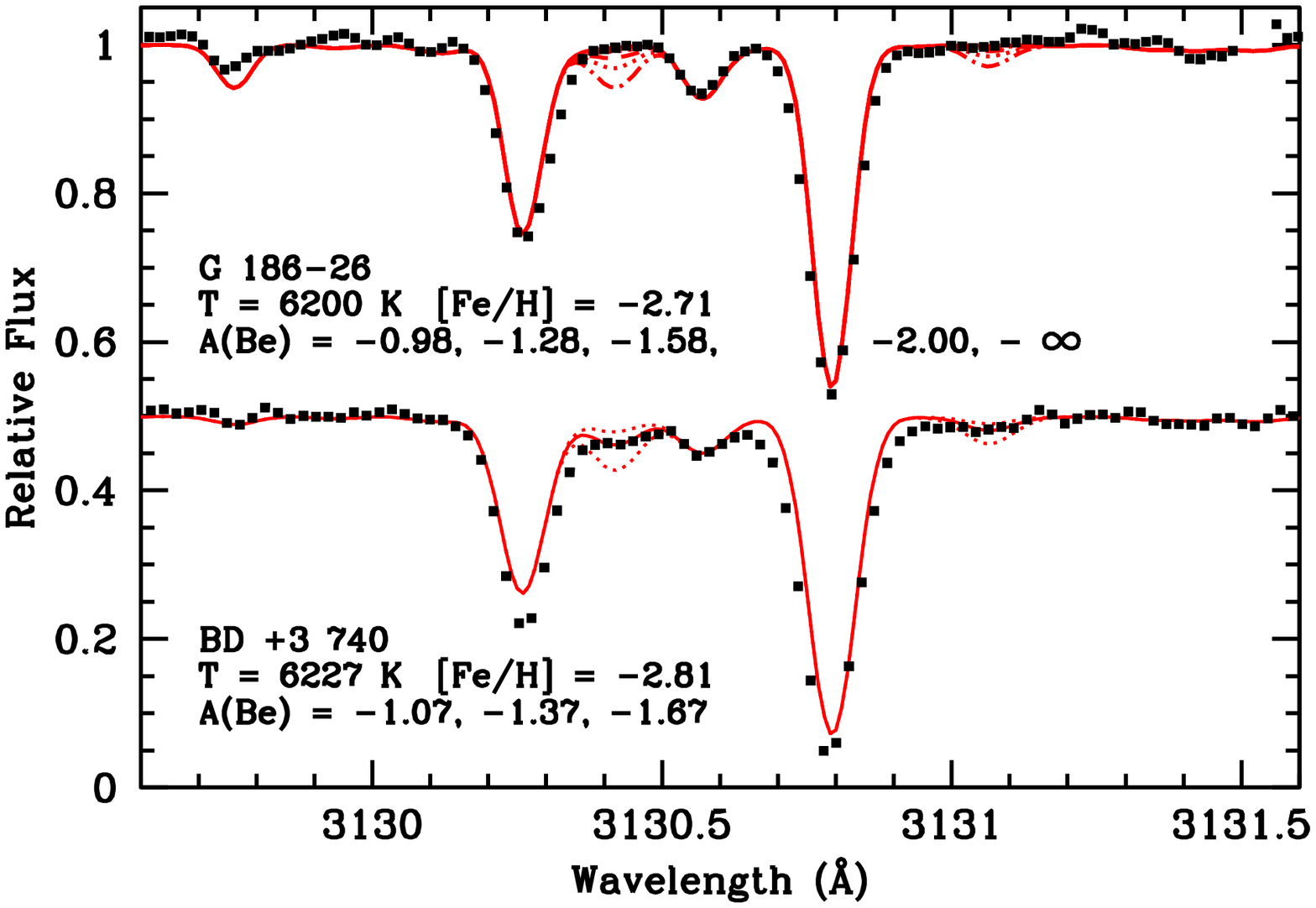} 
\caption{The synthetic spectrum in the Be region for G 186-26.  The small
squares are the data.  The solid line corresponds to our line list predictions
for the Kurucz model for this star for no Be.  The short dashed line
corresponds to the Be abundance expected from the PDD92 calculations ($-$1.58)
and the dotted line is a factor of 2 more Be and the dashed dotted is a factor
of 4 more Be.  The upper dotted line is our upper limit, A(Be) = $-$2.00, but
see Figure 2 for more clarity.  The lower part of the figure is a spectrum of
a star with normal Li and normal Be, BD +3 740 (displaced, but on the same
scale).  G 186-26 and BD +3 740 have similar temperatures and metallicities.
For such metal-poor stars the stronger Be II line $\lambda$3130 is a better
abundance indicator than $\lambda$3131.}
\end{figure}

\begin{figure} 
\plotone{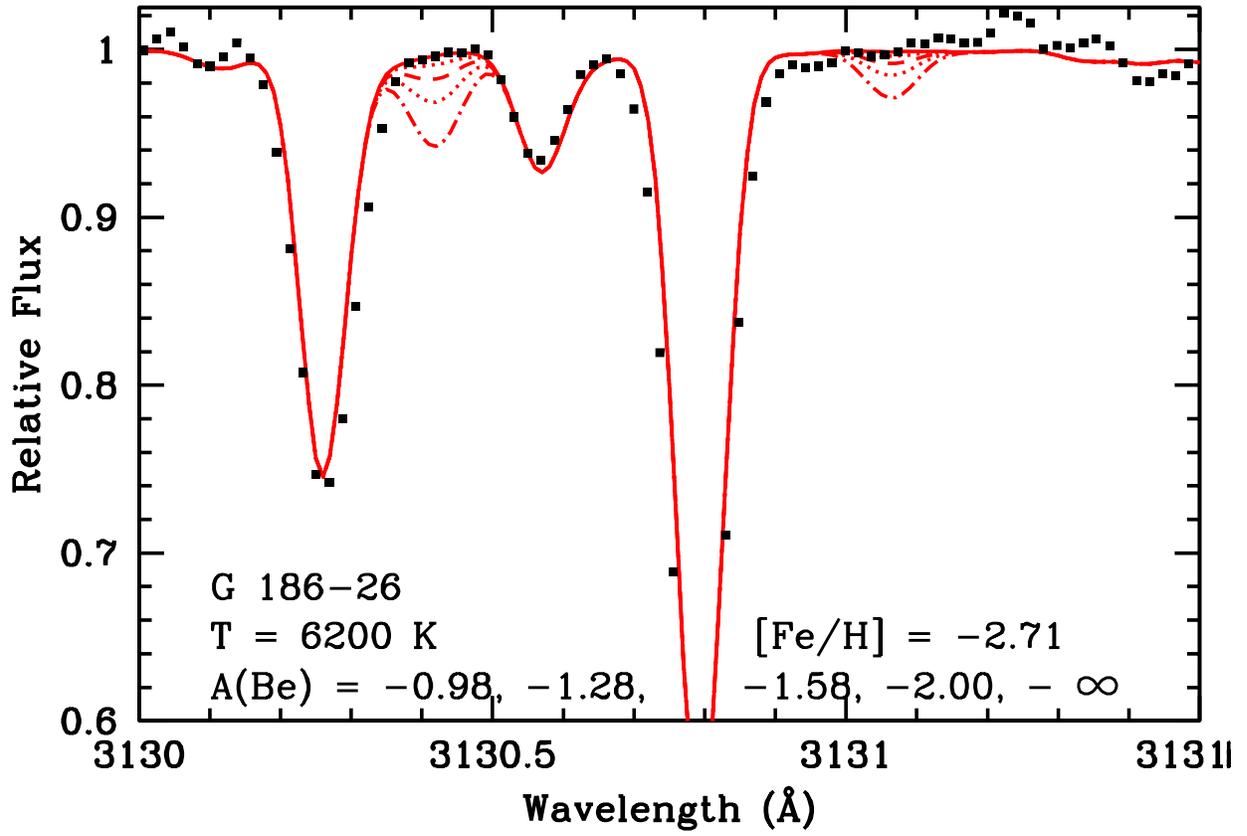} 
\caption{Like Figure 1 of G 186-26, but zoomed in on the Be II lines.}
\end{figure}

\begin{figure} 
\plotone{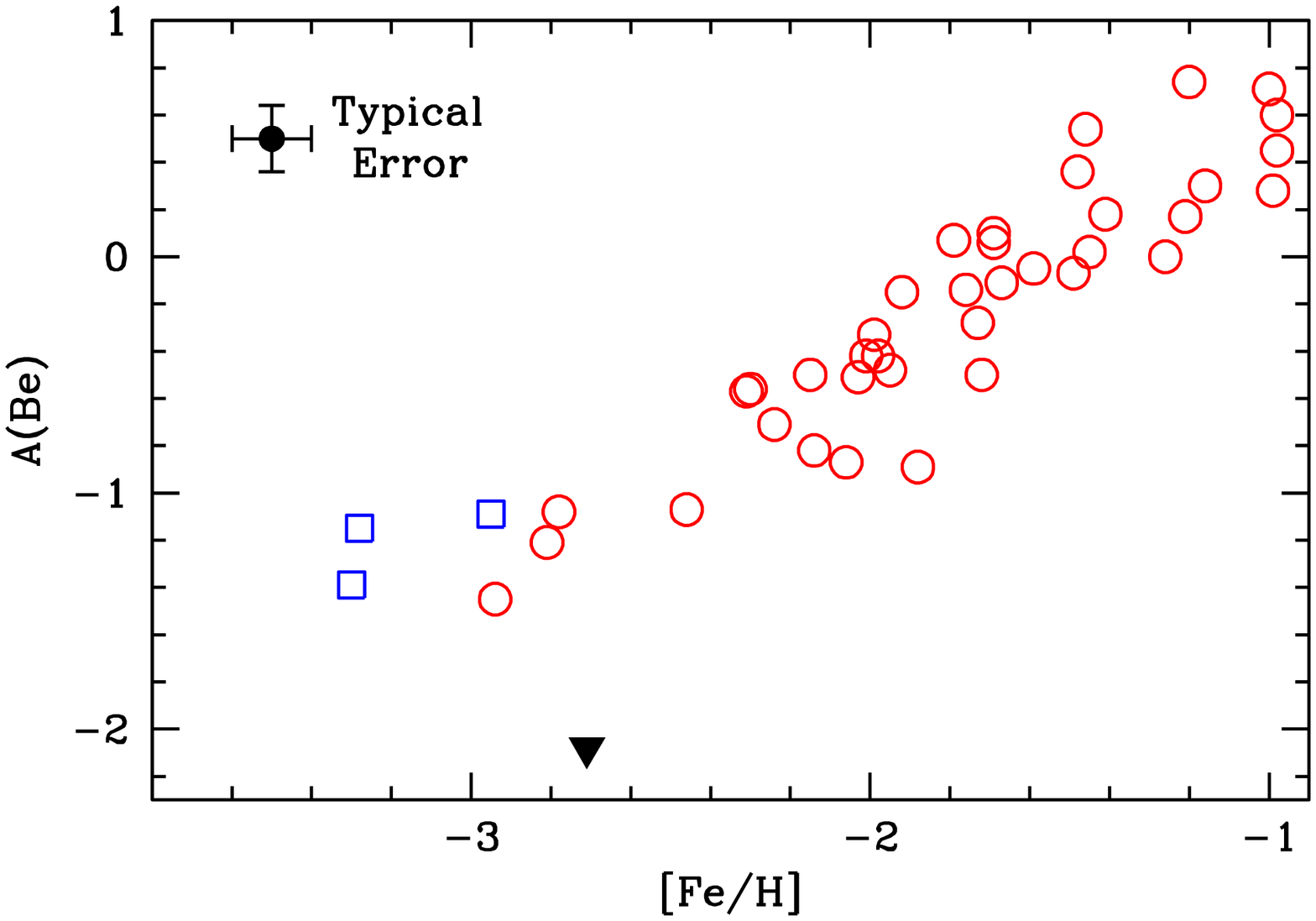} 
\caption{The trend of Fe and Be showing the Be upper limit for G 186-26 as the
solid triangle with Li-normal stars.  Open squares are from Primas et
al.~(2000a, 2000b); open circles are from Boesgaard et al.~(1999), Boesgaard
(2000), Stephens et al.~(1997).}
\end{figure}

\begin{figure} 
\plotone{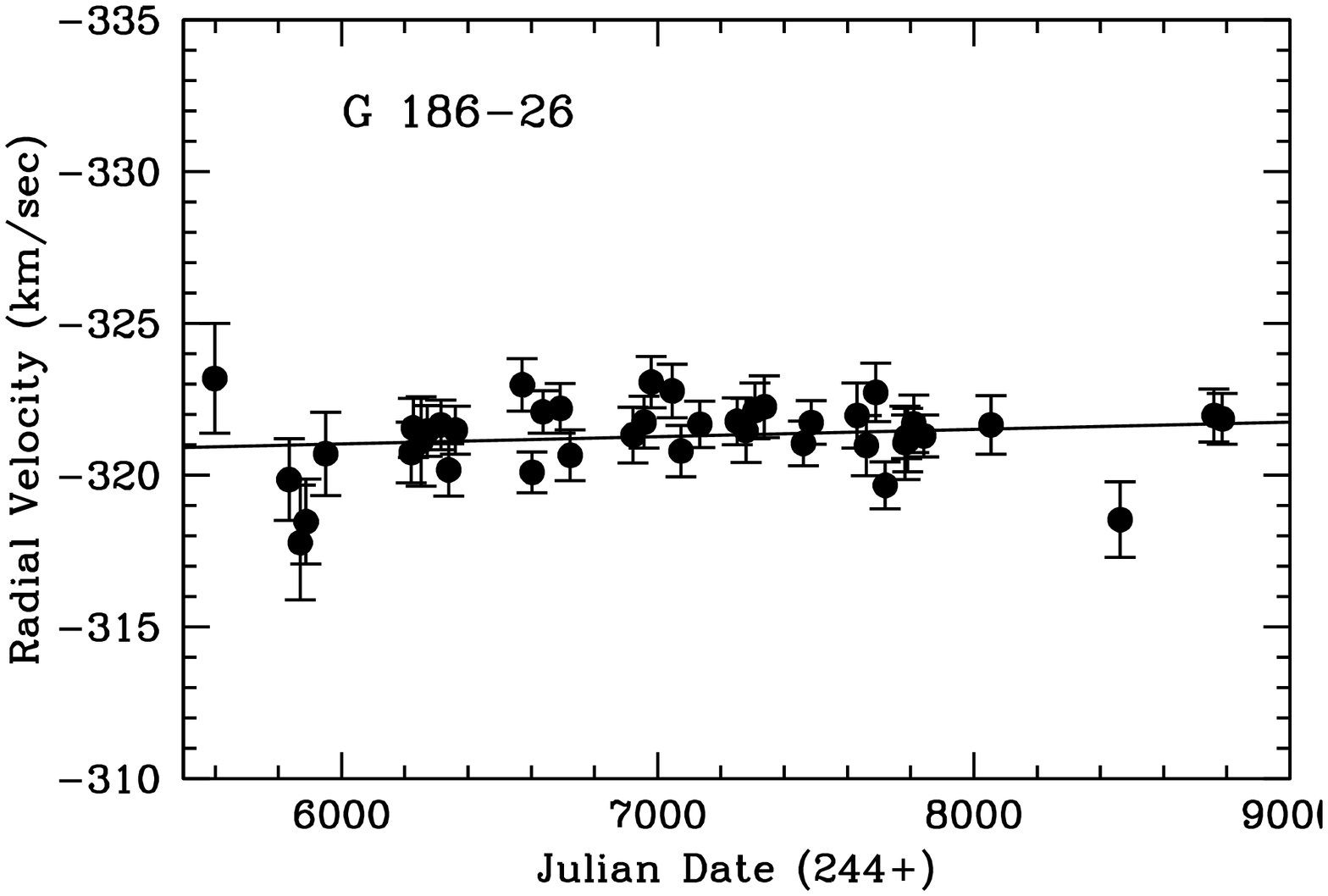} 
\caption{Radial velocities measured for G 186-26 by Carney et al.~(1994).
Latham et al.~2002 find no evidence for binarity in the star.  The line is
simply the least squares fit through the data and is not meant to imply that
there is an increase in the radial velocity over time.}
\end{figure}


\begin{references}

\reference{}
Akerman, C.J., Carigi, L., Nissen, P.E., Pettini, M. \& Asplund, M. 2004,
\aap, 414, 931

\reference{}
Alonso, A., Arribas, S. \& Mart\'\i nez-Roger, C. 1994, \aaps, 107, 365

\reference{}
Alonso, A., Arribas, S. \& Mart\'\i nez-Roger, C. 1996, \aaps, 117, 227

\reference{} Boesgaard, A.M. 2000, in ``The Light Elements and their
Evolution'' IAU Sym. \# 198, ed. L. Da Silva, M. Spite, J.R. Medeiros, p. 389

\reference{}
Boesgaard, A.M. 2004, in ``The Origin and Evolution of the Elements''
ed. A. McWilliam \& M. Rauch, (Cambridge: Cambridge University Press), p. 117

\reference{}
Boesgaard, A.M., Deliyannis, C.P., King, J.R., Ryan, S.G., Vogt, S.S. \&
Beers, T.C. 1999, \aj, 117, 1549 

\reference{}
Bonifacio, P. \& Molaro, P. 1997, \mnras, 285, 847

\reference{}
Carney, B.W. 1983, \aj, 88, 623

\reference{}
Carney, B.W., Latham, D.W. \& Laird, J.B. 2005, \aj, 129, 466

\reference{}
Carney, B.W., Latham, D.W., Laird, J.B. \& Aguilar, L.A. 1994, \aj, 107, 2240

\reference{}
Carney, B.W., Latham, D.W., Laird, J.B., Grant, C.E. \& Morse, J.A. 2001. \aj,
122, 3419

\reference{}
Eggen, O.J. 1979, \apj, 229, 158

\reference{}
Elliott, L. \& Ryan, S.G. 2005, IAU Symposium 228, in press

\reference{}
Fulbright, J. 2002, \aj, 123, 404

\reference{}
Hobbs, L.M. \& Mathieu, R.D. 1991, \pasp, 103, 431

\reference{}
Hobbs, L.M., Welty, D.E. \& Thorburn, J.A. 1991, \apj, 373, L47

\reference{} 
Kurucz, R. 1993, CD-ROM 1, Atomic Data for Opacity Calculations
(Cambridge: SAO)

\reference{}
Latham, D.W., Stefanik, R.P., Torres, G., Davis, R.J., Mazeh, T., Carney,
B.W., Laird, J.B. \& Morse, J.A. 2002, \aj, 124, 1144

\reference{}
Mel\'endez, J. \& Ram\'\i rez, I. 2004, \apj, 615, L33

\reference{}
Nissen, P.E., Gustafsson, B., Edvardsson, B. \& Gilmore, G. 1994, \aap, 285,
440 

\reference{}
Noguchi, K., et al. 2002, \pasj, 54, 855 

\reference{}
Norris, J.E., Ryan, S.G., Beers, T.C. \& Deliyannis, C.P. 1997, \apj, 485, 370

\reference{}
Norris, J.E., Ryan, S.G., \& Beers, T.C. 2001, \apj, 561, 1034

\reference{}
Novicki, M. 2005, Ph.D. Thesis, Univ. of Hawaii at Manoa

\reference{}
Pinsoneault, M. H., Deliyannis, C.P. \& Demarque, P., 1992, \apjs, 78, 179

\reference{}
Pinsoneault, M. H., Steigman, G., Walker, T.P. \& Narayanan, V.K. 2002, \apj,
574, 398

\reference{}
Pinsoneault, M. H., Walker, T.P , Steigman, G. \& Narayanan, V.K. 1999, \apj,
527, 180

\reference{}
Preston, G.W. \& Sneden, C. 2000, \aj, 120, 1014

\reference{}
Primas, F., Asplund, M, Nissen, P.E. \& Hill, V. 2000a, \aap, 364, L42

\reference{}
Primas, F., Asplund, M, Bonifacio, P. \& Hill, V. 2000b, \aap, 362, 666

\reference{}
Pritchet, C.J. \& Glaspey, J.W. 1991, \apj, 373, 105

\reference{}
Ryan, S.G., Norris, J.E. \& Beers, T.C. 1999, \apj, 523, 654

\reference{}
Ryan, S.G., Beers, T.C., Olive, K.A., Fields, B.D. \& Norris, J.E. 2000, \apj,
530, L57

\reference{}
Ryan, S.G., Beers, T.C., Kajino, T. \& Rosolankova, K 2001a, \apj, 547, 231

\reference{} 
Ryan, S.G., Kajino, T., Beers, T.C., Suzuki, T., Romano, D., Matteucci, F. \&
Rosolankova, K 2001b, \apj,

\reference{}
Ryan, S.G., Gregory, S.G., Kolb, U., Beers, T.C. \& Kajino, T. 2002, \apj,
571,  501

\reference{}
Shuster, W.J. \& Nissen, P.E. 1988, \aaps, 73, 225

\reference{}
Sneden, C. 1973, Ph.D. thesis, Univ. of Texas at Austin

\reference{}
Spite, F. \& Spite, M. 1982, \aap, 115, 357

\reference{}
Spite, M., Maillard, J.P. \& Spite, F. 1984, \aap, 141, 56

\reference{}
Spite, M., Molaro, P., Francois, P. \& Spite, F. 1993, \aap, 271, L1

\reference{}
Stephens, A. \& Boesgaard, A.M. 2002, \aj, 123, 1647

\reference{}
Stephens, A., Boesgaard, A.M., King, J.R. \& Deliyannis, C.P. 1997, \apj, 491,
339 

\reference{}
Thorburn, J.A. 1992, \apj, 399, L83

\reference{}
Thorburn, J.A. 1994, \apj, 421, 318

\reference{}
Yi, S.K., Demarque, P. \& Kim, Y. 2004, \apss, 291, 261

\end{references}
\end{document}